\documentclass[10pt,preprintnumbers,twocolumn,nofootinbib,floatfix]{revtex4}

\usepackage{epsfig}
\usepackage{amssymb}
\usepackage{amsmath}
\usepackage{mathrsfs}

\makeindex \textheight=9.5in \textwidth=6.5in \topmargin=0in
\begin{document}
\title{Multiple Unfoldings of Orbifold Singularities:\\Engineering Geometric Analogies to Unification}
\author{Jacob L. Bourjaily}
\email{jbourjai@princeton.edu}
\affiliation{Joseph Henry Laboratories, Princeton University, Princeton, NJ 08544}
\date{2$^{\mathrm{nd}}$ April 2007}

\begin{abstract}
Katz and Vafa \cite{Katz:1996xe} showed how charged matter can arise geometrically by the deformation of ADE-type orbifold singularities in type IIa, M-theory, and F-theory compactifications. In this paper we use those same basic ingredients, used there to geometrically engineer specific matter representations, here to deform the compactification manifold itself in a way which naturally compliments many features of unified model building. We realize this idea explicitly by deforming a manifold engineered to give rise to an $SU_5$ grand unified model into a one giving rise to the Standard Model. In this framework, the relative local positions of the singularities giving rise to Standard Model fields are specified in terms of the values of a small number of complex structure moduli which deform the original manifold, greatly reducing the arbitrariness of their relative positions.
\end{abstract}

\maketitle
\section{Introduction}\vspace{-0.35cm}
One of the ways in which a gauge theory with massless charged matter can arise in type IIa, M-theory, and F-theory is known as geometrical engineering. In this framework, gauge theory at low energy arises from co-dimension four singular surfaces in the compactification manifold \cite{Klemm:1995tj} and charged matter arises as isolated points (curves in F-theory) on these surfaces over which the singularity is enhanced. Katz and Vafa \cite{Katz:1996xe} constructed explicit examples of local geometry which would give rise to different representations of various gauge groups. Their work was presented explicitly in the language of type IIa or F-theory, but the general results have been shown to apply much more broadly to M-theory as well \cite{Atiyah:2001qf,Witten:2001uq,Acharya:2001gy,Acharya:2004qe,Berglund:2002hw}.

The picture of matter and gauge theory arising from pure geometry via singular structures has been used very fruitfully in much of the progress of M-theory phenomenology. In \cite{Witten:2001bf} Witten engineered an interesting phenomenological model in M-theory which could possibly solve the Higgs doublet-triplet splitting problem; this model was explored in great detail together with Friedmann in \cite{Friedmann:2002ty}. There, the explicit topology of the ADE-singular surface and the relative locations of all the isolated conical singularities was motivated by phenomenology---the description of the geometry of the singularities themselves was taken for granted.

Unlike model building with D-branes, for example, geometrical engineering as it has been understood provides little information about the number, type, and relative locations of the many different singularities needed for any phenomenological model. This information must either come {\it a posteriori} from phenomenological success or via duality to a concrete string model. But recent successes in M-theory model building (for example, \cite{Acharya:2006ia,Acharya:2007rc}) motivate a new look at how to describe the relative structure of singularities---at least locally---within the framework of geometrical engineering itself.

In this paper, we reduce the apparent arbitrariness of the number and relative positions of the singularities required by low-energy phenomenology by showing how they can be obtained from deforming a smaller number of singularities in a more unified model. In section \ref{geo} we review the ingredients of geometrical engineering as described in \cite{Katz:1996xe}. The basic framework is then interpreted in a novel way in section \ref{modelbuilding} to relate manifolds with matter singularities to those with more or less symmetry. The idea is used explicitly to deform an $SU_5$ grand unified model into the Standard model.

To be clear, as in \cite{Katz:1996xe} our results apply only strictly to $\mathscr{N}=2$ models from type IIa compactifications or $\mathscr{N}=1$ models from F-theory compactifications\footnote{We essentially describe non-compact Calabi-Yau three-folds which are $K3$-fibrations over $\mathbb{C}^1$. If the $\mathbb{C}^1$-base is fibred over $\mathbb{CP}^1$ as an $\mathscr{O}(-2)$ bundle, for example, then the total space will be a Calabi-Yau four-fold upon which F-theory can be compactified, giving rise to an $\mathscr{N}=1$ theory.}; but we suspect that this framework has an M-theory analogue in the spirit of \cite{Berglund:2002hw}.
\vspace{-0.35cm}
\section{Geometrical Engineering\label{geo}}\vspace{-0.35cm}
In the framework of geometrical engineering the compactification manifold is described as a fibration of (singular) $K3$ surfaces over a base space of appropriate dimension. The collection of point-like (co-dimension four) singularities of the $K3$ fibres would then be a co-dimension four surface in the compactified manifold, giving rise to gauge theory of type corresponding to the singularities on each $K3$ fibre. Table \ref{orbifolds} lists polynomials in $\mathbb{C}^3$ whose solutions can be (locally) taken to be the fibres for each corresponding gauge group.

One of the strengths of this framework is its generality: the local geometry is specified in terms of the $K3$ fibres, so that the description applies equally well to compactifications in type IIa, M-theory, and F-theory---the difference being the dimension of the space over which the surfaces in Table \ref{orbifolds} are fibred.

\begin{table}[t]\caption{\label{orbifolds}Hypersurfaces in $\mathbb{C}^3$ giving rise to the desired orbifold singularities.}
\begin{tabular}{lr}
Gauge group&Polynomial\\
\hline $SU_n$ ($\equiv A_{n-1}$) & $xy=z^n$\\
$SO_{2n}$ ($\equiv D_{n}$) & $x^2+y^2z=z^{n-1}$\\
$E_6$ & $x^2=y^3+z^4$\\
$E_7$ & $x^2+y^3=yz^3$\\
$E_8$ & $x^2+y^3=z^5$
\end{tabular}\end{table}

To obtain massless charged matter, however, additional structure is necessary. Specifically, at isolated points (in type IIa or M-theory) on the co-dimension four singular surface, the type of singularity of the $K3$ fibres must be enhanced by one rank. Mathematically, this requires that one can describe how the various polynomials in Table \ref{orbifolds} can be deformed into each other; and the possible two-dimensional deformations have been classified \cite{Katz:1992ab}.

For example, to describe the embedding of a massless $\mathbf{5}$ of $SU_5$ in type IIa, you would need to start with a $K3$-fibred Calabi-Yau where each of the fibres are of the type giving rise to $SU_5$ gauge theory. From Table \ref{orbifolds} we see that these four-dimensional fibres are locally the set of solutions to the equation \begin{equation}xy=z^5,\end{equation} in $\mathbb{C}^3$. Now, to obtain matter in the $\mathbf{5}$ representation, there would need to be an isolated point somewhere on the two-dimensional base space where the fibre is enhanced to $SU_6$, \cite{Katz:1996xe}. A description of the local geometry can be given by \begin{equation}xy=(z+5t)(z-t)^5,\label{su6tosu5}\end{equation} where $t$ is a complex coordinate on the base over which the $K3$'s are fibred. Notice that when \mbox{$t=0$} the equation describes precisely the fibre which would have given rise to $SU_6$ gauge theory if it were fibred over the entire base manifold. However, because it is the fibre only over the origin in the complex $t$-plane, there is no $SU_6$ gauge theory. Equation (\ref{su6tosu5}) is said to describe the `resolution' $SU_6\to SU_5$, which is found to give rise to $SU_5$ gauge theory at low energy with a single massless $\mathbf{5}$ at $t=0$. This and many other explicit examples of such resolutions and the matter representations obtained are given in \cite{Katz:1996xe}.

One subtlety which makes the description above not automatically apply to M-theory constructions, however, is that in equation (\ref{su6tosu5}) the complex parameter $t$ is two-dimensional: taken as a coordinate over which the $K3$ surfaces are fibred, it gives rise to a six-dimensional compactification manifold. In M-theory, co-dimension four singularities are three-dimensional and chiral matter would live at isolated points on these three dimensional orbifold singularities. So in M-theory the resolution \mbox{$SU_6\to SU_5$} would need a three-dimensional deformation. Morally, the structure is identical to that described in equation (\ref{su6tosu5}), but the parameter $t$ must be upgraded to describe three-dimensional deformations. This can be done in terms of hyper-K\"{a}hler quotients. We suspect that all the resolutions described explicitly for type IIa here and in \cite{Katz:1996xe} can be upgraded to three-dimensional deformations needed in M-theory, and in many cases these generalizations have already been given \cite{Witten:2001uq,Acharya:2001gy,Berglund:2002hw}.

\section{Engineering Geometric Analogies to Unification\label{modelbuilding}}\vspace{-0.35cm}
The main result of this paper is that distinct conical singularities on a surface with some gauge symmetry can be deformed into each other in ways analogous to unification; and conversely, that a description of a single matter field in a unified theory can be `unfolded' into distinct matter fields in a theory of lower gauge symmetry. Because the tools used to perform these unification-like deformations are precisely the same as those used to describe the singularities themselves, some care must be taken to avoid unnecessary confusion.

We will start by reinterpreting the tools used above to engineer charged matter, and then we will use both interpretations simultaneously to construct explicit examples of the geometric analogue to unified model building.

Consider again the resolution $SU_6\to SU_5$ described by \begin{equation}xy=(z+5s)(z-s)^5,\label{gsu6tosu5}\end{equation} where we have replaced $t\mapsto s$ from equation (\ref{su6tosu5}) to make a interpretative distinction that will soon become clear. We propose to momentarily discuss pure gauge theory and ignore any description of matter. With this in mind, take a fixed (real) two-dimensional neighborhood over which {\it every} point is fibred by the solutions to equation (\ref{gsu6tosu5}) for any fixed value of $s$. Because the fibres are the same everywhere on the manifold, there is no matter: for any $s$ the geometry would give rise to pure gauge theory at low energy. For $s\neq0$ solutions to equation (\ref{gsu6tosu5}) are $SU_5$ fibres and so the compactification manifold would give rise to pure $SU_5$. However, when $s=0$ the fibres are all $SU_6$ and so the low-energy theory would be pure $SU_6$. Therefore $s$ is a `global' parameter which deforms the gauge content of the theory such that for arbitrary values of $s\neq0$ the theory is pure $SU_5$ and for $s=0$ it is pure $SU_6$. That this deformation is `smooth' is apparent at least when $s\neq0$.

An obvious question to ask is how this framework applies when conical singularities are present. We will show that when the ADE-surface singularity changes because of some complex structure modulus such as $s$ above, the conical singularities giving rise to charged matter (often) behave as one would expect from unified model building intuition. This is best demonstrated with explicit examples.

Suppose that the singular $K3$ surfaces are fibred over a two-dimensional base space with local complex coordinate $t$. And say the four-dimensional fibre over the point $t$ is given by the solutions to \begin{equation}xy=(z+5t)(z-t+3s)^2(z-t-2s)^3,\label{5tosm}\end{equation} for a given value of $s$, which is now to be interpreted as a complex structure modulus deforming the entire local geometry near $t=0$. When $s=0$ the geometry is of course identical to our previous description of $SU_6\to SU_5$ and so the theory would be $SU_5$ with a single massless $\mathbf{5}$ located at $t=0$.

Consider now $s$ to be fixed at some non-zero value. The gauge theory is then $SU_3\times SU_2\times U_1$: for generic values of $t$, the fibres given by equation (\ref{5tosm}) have two singular points, at \mbox{$x=y=z-t+3s=0$} and \mbox{$x=y=z-t-2s=0$}, and so the union of these points over the base manifold coordinatized by $t$ will be two distinct, two-dimensional singular surfaces: one giving rise to $SU_2$ and the other $SU_3$. These surfaces become coincident as a single $SU_5$ surface when $s=0$.
\epsfxsize=2.4in
\epsfysize=2.4in
\begin{figure}[t]\includegraphics[scale=1]{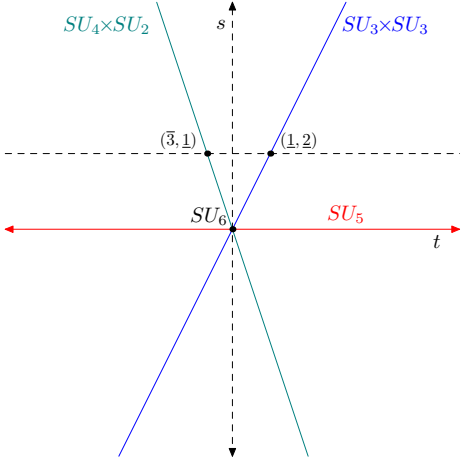}
\caption{The $t$-$s$ plane describing the deformation of a theory with a single $\mathbf{5}$ of $SU_5$ into one of \mbox{$SU_3\times SU_2\times U_1$} gauge theory with one $(\overline{\mathbf{3}},\mathbf{1})_{1/3}$ and one $(\mathbf{1},\mathbf{2})_{-1/2}$ as a function of $s$ as described by equation (\ref{5tosm}). For a fixed value of $s$, the base space over which solutions to (\ref{5tosm}) are fibred are indicated by the black line. Notice that the relative positions of the two isolated (conical) singularities are fixed by $s$.\label{su6doubleres}}\end{figure}

Along the complex $t$-plane, there are two isolated points over which the singularities are enhanced: at \mbox{$t=s/2$} the fibre is visibly $SU_3\times SU_3$, and at \mbox{$t=-s/3$} the fibre is $SU_4\times SU_2$. Therefore the theory has two two charged, massless fields, in the $(\mathbf{1},\mathbf{2})_{-1/2}$ and $(\overline{\mathbf{3}},\mathbf{1})_{1/3}$ representations of $SU_3\times SU_2\times U_1$ at $t=s/2$ and $t=-s/3$, respectively. Figure \ref{su6doubleres} indicates the singularity structure as a function of $s$.

Notice how this description parallels unified model building: the $s=0$ theory of one $\mathbf{5}$ of $SU_5$ deforms smoothly into one with $(\overline{\mathbf{3}},\mathbf{1})_{1/3}\oplus(\mathbf{1},\mathbf{2})_{-1/2}$ of \mbox{$SU_3\times SU_2\times U_1$.}

Similarly, we may ask how a $\mathbf{10}$ of $SU_5$ would deform into distinct singularities supporting Standard Model matter fields. The fibre structure giving rise to a massless $\mathbf{10}$ of $SU_5$ is given as follows. Let $t$ be a local coordinate on the base space over which fibres are given by solutions to \begin{equation}x^2+y^2z+2yt^5=\frac{1}{z}\left\{\left(z+t^2\right)^5-t^{10}\right\};\label{so10tosu5}\end{equation} at $t=0$, equation (\ref{so10tosu5}) describes an $SO_{10}$ fibre, while for $t\neq0$ the fibres are $SU_5$---although in this case the result is harder to read off. This resolution, $SO_{10}\to SU_5$, gives rise to a $\mathbf{10}$ of $SU_5$ \cite{Katz:1996xe}.

\begin{figure}[t]\includegraphics[scale=1]{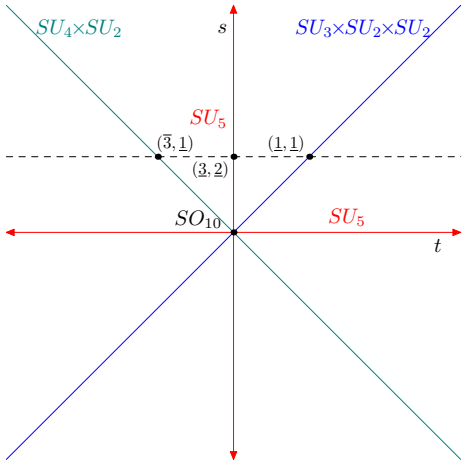}
\caption{The $t$-$s$ plane describing the deformation of a theory with a single $\mathbf{10}$ of $SU_5$ into one of $SU_3\times SU_2\times U_1$ gauge theory, with matter content $(\overline{\mathbf{3}},\mathbf{1})_{-2/3}\oplus(\mathbf{3},\mathbf{2})_{1/6}\oplus(\mathbf{1},\mathbf{1})_{1}$, as a function of $s$ as described by equation (\ref{so10double}). For a fixed value of $s$, the base space over which solutions to (\ref{so10double}) are fibred are indicated by the black line. Notice that the relative positions of the three isolated (conical) singularities are fixed by $s$.\label{so10doubleres}}\end{figure}

Following the same idea as before, the deformation of this singularity into $SU_3\times SU_2$ is given by \begin{widetext}\begin{equation}x^2+y^2z+2y(t+s)^3(t-s)^2=
\frac{1}{z}\left\{\left(z+(t-s)^2\right)^2\left(z+(t+s)^2\right)^3-(t-s)^4(t+s)^6\right\},\label{so10double}\end{equation}\end{widetext} where $s$ is again interpreted as a complex structure modulus deforming the geometry near the singularity. Notice as before that $s=0$ describes an $SU_5$ theory with one massless $\mathbf{10}$ located at $t=0$. However, for $s\neq0$ there are again two orbifold singularities corresponding to $SU_3\times SU_2\times U_1$ gauge theory. At three distinct points on the complex $t$ plane the rank of the fibre is enhanced: $t=-s$, $t=0$, and $t=s$ give rise to matter in the $(\overline{\mathbf{3}},\mathbf{1})_{-2/3}$, $(\mathbf{3},\mathbf{2})_{1/6}$, and $(\mathbf{1},\mathbf{1})_{1}$ representations of $SU_3\times SU_2\times U_1$, respectively. The structure of the deformation achieved by varying $s$ is shown in Figure \ref{so10doubleres}.

Again, our intuition from unified model building \nopagebreak[4]is realized naturally in this framework.\newpage

\vspace{-0.35cm}\section{Discussion}\vspace{-0.35cm}
One of the primary reasons why geometrical engineering had not been more widely used phenomenologically is because the number, type, and relative locations of the singularities giving rise to various matter fields were explicitly {\it ad hoc}: the inherent local framework prevented relationships between distinct singularities from being discussed.  In this paper, we have shown a framework in which these questions can be addressed concretely, systematically reducing the arbitrariness of these models.

Of course, the local nature of geometrical engineering is still inherent in this framework, and continues to prevent us from addressing questions about the global structure such as stability, quantum gravity, and the quantization of seemingly continuous parameters like $s$. However, in the spirit of  \cite{Verlinde:2005jr}, we think that local engineering is a good step toward realistic string phenomenology, and may perhaps offer new insights.

In this paper we explicitly illustrated the geometric unfolding of the matter content of an $SU_5$ grand unified model into the Standard Model. But the procedure can easily be generalized. It is not difficult to see how this will work for a more unified theory. For example, one can envision how an entire family could unfold out of a single $E_6\to SO_{10}$ resolution (which starts as a $\mathbf{16}$ of $SO_{10}$), or how all three families of the Standard Mode could be unfolded out of a single $E_8\to SO_{10}\times SU_3$ or $E_8\to E_6\times SU_2$ resolution. However, these examples require more sophisticated tools of analysis, and so we have chosen to describe these in a forthcoming work.

\section{Acknowledgements}\vspace{-0.5cm}
This work originated from discussions with Malcolm Perry whose insights drove this work forward in its earliest steps. The author also appreciates helpful discussions, comments, and suggestions from Herman Verlinde, Sergei Gukov, Gordon Kane, Edward Witten, Paul Langacker, Bobby Acharya, Dmitry Malyshev, Matthew Buican, Piyush Kumar, and Konstantin Bobkov.

This research was supported in part by the Michigan Center for Theoretical Physics and a Graduate Research Fellowship from the National Science Foundation.


\begin{thebibliography}{10}

\bibitem{Katz:1996xe}
S.~Katz and C.~Vafa, ``Matter from {G}eometry,'' {\em Nucl. Phys.}, vol.~B497,
  pp.~146--154, 1997, hep-th/9606086.

\bibitem{Klemm:1995tj}
A.~Klemm, W.~Lerche, and P.~Mayr, ``K3 {F}ibrations and {H}eterotic {T}ype
  {I}{I} {S}tring {D}uality,'' {\em Phys. Lett.}, vol.~B357, pp.~313--322,
  1995, hep-th/9506112.

\bibitem{Atiyah:2001qf}
M.~Atiyah and E.~Witten, ``M-theory {D}ynamics on a {M}anifold of ${G}_2$
  {H}olonomy,'' {\em Adv. Theor. Math. Phys.}, vol.~6, pp.~1--106, 2003,
  hep-th/0107177.

\bibitem{Witten:2001uq}
E.~Witten, ``Anomaly {C}ancellation on ${G}_2$ {M}anifolds,'' 2001,
  hep-th/0108165.

\bibitem{Acharya:2001gy}
B.~Acharya and E.~Witten, ``Chiral {F}ermions from {M}anifolds of ${G}_2$
  {H}olonomy,'' 2001, hep-th/0109152.

\bibitem{Acharya:2004qe}
B.~S. Acharya and S.~Gukov, ``M-theory and {S}ingularities of {E}xceptional
  {H}olonomy {M}anifolds,'' {\em Phys. Rept.}, vol.~392, pp.~121--189, 2004,
  hep-th/0409191.

\bibitem{Berglund:2002hw}
P.~Berglund and A.~Brandhuber, ``Matter from ${G}_2$ {M}anifolds,'' {\em Nucl.
  Phys.}, vol.~B641, pp.~351--375, 2002, hep-th/0205184.

\bibitem{Witten:2001bf}
E.~Witten, ``Deconstruction, {$G_2$} {H}olonomy, and {D}oublet-{T}riplet
  {S}plitting,'' 2001, hep-ph/0201018.

\bibitem{Friedmann:2002ty}
T.~Friedmann and E.~Witten, ``Unification {S}cale, {P}roton {D}ecay, and
  {M}anifolds of {G}(2) {H}olonomy,'' {\em Adv. Theor. Math. Phys.}, vol.~7,
  pp.~577--617, 2003, hep-th/0211269.

\bibitem{Acharya:2006ia}
B.~Acharya, K.~Bobkov, G.~Kane, P.~Kumar, and D.~Vaman, ``An {M} {T}heory
  {S}olution to the {H}ierarchy {P}roblem,'' {\em Phys. Rev. Lett.}, vol.~97,
  p.~191601, 2006, hep-th/0606262.

\bibitem{Acharya:2007rc}
B.~S. Acharya, K.~Bobkov, G.~L. Kane, P.~Kumar, and J.~Shao, ``Explaining the
  {E}lectroweak {S}cale and {S}tabilizing {M}oduli in {M}-{T}heory,'' 2007,
  hep-th/0701034.

\bibitem{Katz:1992ab}
S.~Katz and D.~Morrison, ``Gorenstein {T}hreefold {S}ingularities with {S}mall
  {R}esolutions via {I}nvariant {T}heory for {W}eyl {G}roups,'' {\em J.
  Algebraic Geometry}, vol.~1, pp.~449--530, 1992.

\bibitem{Verlinde:2005jr}
H.~Verlinde and M.~Wijnholt, ``Building the {S}tandard {M}odel on a
  {D}3-{B}rane,'' {\em JHEP}, vol.~01, p.~106, 2007, hep-th/0508089.

\end{thebibliography}

\end{document}